# Effect of clustering on ellipsometric spectra of randomly distributed gold nanoparticles on a substrate

**Huai-Yi Xie,[1] Yia-Chung Chang,[1,*] Guangwei Li,[1] and Shih-Hsin Hsu[1,2]**

[1]*Research Center for Applied Sciences, Academia Sinica, 128 Academia Road, Sec. 2, Taipei 11529, Taiwan*

[2] *Centre for Ultrahigh bandwidth Devices for Optical Systems (CUDOS), Institute of Photonics and Optical Science (IPOS), School of Physics, University of Sydney, NSW 2006, Australia*

## Abstract

We present a theoretical model for describing light scattering from randomly distributed Au nanoparticles on a substrate, including the clustering effect. By using the finite-element Green's function method and spherical harmonic basis functions, we are able to calculate the polarization-dependent reflectivity spectra of the system (modeled by randomly distributed nanoparticles coupled with clusters) efficiently and accurately. The calculated ellipsometric spectra of the system with clusters can adequately describe the experimental data for the whole frequency range. We find that the clustering effect leads to some prominent features in the low frequency range of the ellipsometric spectra, which are attributed to plasmonic resonances associated with the coupling of Au nanoparticles and clusters.

[*]Corresponding author: yiachang@gate.sinica.edu.tw

## I. Introduction

Recently, optical properties of nanoparticles have been widely studied by scientists, especially in phenomena concerning enhancement of electromagnetic waves near the surface of nanoparticles due to surface plasmon resonance (SPR). The behavior of SPR is very sensitive to the shape and dielectric constants of media surrounding the nanoparticles. [1] SPR is wildly used in many scientific studies such as surface enhanced Raman scattering (SERS) [2-7] and spectroscopic ellipsometry (SE) measurements. [8-11]

Theoretically, the scattering from an isolated sphere has been analytically studied from the conventional Mie scattering theory which is based on vector spherical harmonic functions and matching boundary conditions at the surface of the sphere. [12-13] Theory for the scattering from an isolated spheroid has been also constructed [14]. Furthermore, many methods were used to investigate the optical properties of an isolated sphere or spheroid on a substrate. However, these methods are based on some approximations such as the image method [15] and electrostatic formulation [16-18]. For light scattering from periodic distribution of nanoparticles on multilayer films, rigorous coupled wave analysis (RCWA) has been adopted. [19, 20] On the other hand, the finite-element Green's function (FEGF) approach [21, 22] has been developed and shown to be more efficient than RCWA except that the FEGF approach sometimes does not converge fast enough for wavelengths near the plasmonic resonance. For random distribution of nanoparticles on multilayer films, the effective medium approximation (EMA) and so-called the GranFilm code [23] are suitable for describing the optical properties in the long wavelength range, while the FEGF method is appropriate for all frequency range since it takes into account the retardation effect. [21, 22] The random distribution of metallic nanoparticles on a substrate may be used as a mask for fabricating nanostructures. [24] Such samples have also been used as biosensing chips. [25, 26] Therefore, it would be desirable to develop a reliable modeling technique in order to perform optical inspection of the morphology of these samples without resorting to the electron diffraction for metrology purposes. However, in preparing these nanoparticle covered samples, it is difficult to avoid the formation of clusters due to aggregation of nanoparticles. To our knowledge, such a clustering effect of nanoparticles on the reflectivity spectra has never been studied theoretically with suitable methods.

Figure 1 shows the scanning electron microscopy (SEM) pictures of the samples considered in Ref. [22]. Random distribution of Au nanoparticles with diameters ranging from 20 nm to 80 nm can be seen in these pictures. For sample (a) d = 20 nm, the distribution is rather uniform with little cluster formation. For the other three samples, the formation of clusters is obvious.

In this paper, we present a theoretical model, which can describe the effect of nanoparticle clusters on the reflectivity spectra with reasonable success. We use a spherical harmonics-based Green's function (SHGF) approach to calculate light scattering from randomly distributed nanoparticles with and without coupling to clusters. In our modeling, the experimental dielectric constants of bulk Au [27] were adopted for the interior of nanoparticles. For the multilayered substrate, the optical contrast between the APTES coating and the glass slide was found to be extremely small. Hence, we can treat the substrate plus APTES as a single semi-infinite layer and use its pseudo-dielectric constants measured experimentally as inputs in our theoretical calculations. A previous version of this study was published in Ref. [28]. The current version represents an update, which corrects a minor error of Ref. [28] and provides convergent results done with a larger set of basis functions.

**II. Modeling for light scattering from randomly distributed nanoparticles with clusters**

The modeling of the system without clustering effect has been briefly described in [20, 22]. Here, we give a more detailed description in order to address the clustering effect. Rather than using the cylindrical basis functions with flexible radial part as adopted in [20, 22], here we use spherical harmonics basis functions with rigid radial part which satisfy the Maxwell equations. This reduces the number of basis functions needed to describe the wave function inside a nearly spherical nanoparticle, and improves the efficiency and accuracy. First, we consider a random distribution of nanoparticles of identical size and shape, with an average separation of $p$ (the pitch). For convenience, we can write the wave function for the electric field in the form of linear combination of localized orbitals (LCAO) (for $\mathbf{r}$ restricted in the layer where the nanoparticles reside)

$$\mathbf{E}(\mathbf{r}) = \frac{1}{\sqrt{N}} \sum_i e^{i\mathbf{k}_0 \cdot \mathbf{R}_i} \mathbf{u}_i(\mathbf{r}_i), \qquad (1)$$

where $\mathbf{k}_0$ denotes the wave vector of the incident wave, $\mathbf{r}_i \equiv \mathbf{r} - \mathbf{R}_i$ denotes the spatial coordinate relative to a scatterer centered at $\mathbf{R}_i$ and $\mathbf{u}_i(\mathbf{r}_i)$ is a local function which is non-zero only for $\mathbf{r}$ in a cell surrounding $\mathbf{R}_i$. $N$ is the number of scatters in the sample area of interest. The Lippmann-Schwinger (L-S) equation [29] for $N$ scatters on a multilayer film reads,

$$e^{i\mathbf{k}_0 \cdot \mathbf{R}_i} \mathbf{u}_i(\mathbf{r}_i) = \sqrt{N} \mathbf{E}_0(\mathbf{r}) + \sum_{j=1}^{N} \int d\mathbf{r}' \mathbf{G}(\mathbf{r}, \mathbf{r}') \cdot V_j(\mathbf{r}') e^{i\mathbf{k}_0 \cdot \mathbf{R}_j} \mathbf{u}_j(\mathbf{r}'_j), \qquad (2)$$

where $\mathbf{r}$ is restricted in cell $i$, $\mathbf{G}(\mathbf{r}, \mathbf{r}')$ denotes the dyadic Green's function (GF)

for the uniform multilayer background material, and $\mathbf{E}_0(\mathbf{r})$ denotes the unperturbed electric field (i.e. solution to the system without the scatterers). $V_j(\mathbf{r})$ describes the perturbation due to replacing the dielectric constant of the background material by the nanoparticle. We define the Fourier expansion of the GF suitable for a laminated structure stacked along the z axis,

$$\mathbf{G}(\mathbf{r},\mathbf{r}') = \frac{1}{(2\pi)^2} \int d\mathbf{k}_n e^{i\mathbf{k}_n \cdot (\boldsymbol{\rho}-\boldsymbol{\rho}')} \mathbf{g}_n(z,z'), \tag{3}$$

where $\mathbf{g}_n(z,z')$ is given in Ref. [21]. Here, $\boldsymbol{\rho}$ ($\boldsymbol{\rho}'$) denotes the projection of $\mathbf{r}$ ($\mathbf{r}'$) in the x-y plane, and $\mathbf{k}_n$ denotes any wave vector in the x-y plane, since the system is not periodic, unlike in Ref. [21]. Substituting Eq. (3) into Eq. (2) yields

$$\mathbf{u}_i(\mathbf{r}_i) = \sqrt{N} e^{-i\mathbf{k}_0 \cdot \mathbf{R}_i} \mathbf{E}_0(\mathbf{r})$$
$$+ \sum_{j=1}^{N} \int \frac{d\phi_n}{(2\pi)^2} \int k_n dk_n \int d\mathbf{r}'_j e^{i\mathbf{k}_n \cdot (\boldsymbol{\rho}_i - \boldsymbol{\rho}'_j)} \mathbf{g}_n(z,z') \cdot e^{-i\mathbf{K}_n \cdot (\mathbf{R}_j - \mathbf{R}_i)} V_j(\mathbf{r}'_j) \mathbf{u}_j(\mathbf{r}'_j), \tag{4}$$

where $\mathbf{K}_n \equiv \mathbf{k}_n - \mathbf{k}_0$, and $\boldsymbol{\rho}_i \equiv \boldsymbol{\rho} - \mathbf{R}_i$ ($\boldsymbol{\rho}_j \equiv \boldsymbol{\rho} - \mathbf{R}_j$). For a random but uniform distribution of identical scatterers with cylindrical symmetry, all local functions $\mathbf{u}_j(\mathbf{r}'_j)$ are almost the same, and the L-S equation reduces to

$$\mathbf{u}_i(\mathbf{r}_i) = \sqrt{N} e^{-i\mathbf{k}_0 \cdot \mathbf{R}_i} \mathbf{E}_0(\mathbf{r})$$
$$+ \int \frac{d\phi_n}{(2\pi)^2} \int k_n dk_n S(\mathbf{K}_n) \int d\mathbf{r}'_i e^{i\mathbf{k}_n \cdot (\boldsymbol{\rho}_i - \boldsymbol{\rho}'_i)} \mathbf{g}_n(z,z') \cdot V_1(\mathbf{r}'_i) \mathbf{u}_i(\mathbf{r}'_i), \tag{5}$$

where $V_j = V_1$ are identical for all scatterers, thus dropping the subscript $j$. $V_1(\mathbf{r}') = (\varepsilon_a - 1)(\omega/c)^2$ for $\mathbf{r}'$ inside the nanoparticle and vanishes otherwise.

$$S(\mathbf{K}_n) = \frac{1}{N} \sum_i \sum_j e^{-i\mathbf{K}_n \cdot (\mathbf{R}_j - \mathbf{R}_i)} \tag{6}$$

is the structure factor. For a period array of identical scatterers, we have

$$S(\mathbf{K}_n) = N \sum_{\mathbf{G}} \delta_{\mathbf{K}_n, \mathbf{G}}, \tag{7}$$

where $\mathbf{G}$ denote the reciprocal lattice vectors in the plane. Thus, our theory can reproduce the results of a periodic grating as described in Ref. [21]. For a random distribution of nearly identical scatters, the structure factor can be written in the form,

$$S(\mathbf{K}_n) = 1 + fS_1(\mathbf{K}_n), \tag{8}$$

where we have introduced a "similarity" factor $f$, which describes the ratio of the average of wave functions over all other sites to the on-site wave function. If wave functions at all sites are identical and the coherence is maintained, we will have $f = 1$ for all wavelengths.

$$S_1(\mathbf{K}_n) = \frac{1}{N}\sum_i \sum_{j \neq i} e^{-i\mathbf{K}_n \cdot (\mathbf{R}_j - \mathbf{R}_i)}. \tag{9}$$

The first term in Eq. (8) represents the on-site scattering process, while the second term is responsible for the off-site scattering process, which can be evaluated approximately by replacing the sum over discrete random sites with a continuous integral. Then, we have

$$\begin{aligned} S_1(\mathbf{K}_n) &= \frac{1}{N}\sum_i \sum_{j \neq i} e^{-i\mathbf{K}_n \cdot (\mathbf{R}_j - \mathbf{R}_i)} \approx \sum_{j \neq 1} e^{-i\mathbf{K}_n \cdot (\mathbf{R}_j - \mathbf{R}_1)} \approx \frac{1}{A_{cell}} \int d\phi \int_{R_u}^{\infty} e^{i\mathbf{K}_n \cdot \mathbf{R}} R dR \\ &= \delta_{\mathbf{k}_n, \mathbf{k}_0} - 2\frac{J_1(K_n R_u)}{K_n R_u} \end{aligned} \tag{10}$$

where a region surrounding the site $\mathbf{R}_1$ with an area $A_{cell} = p^2$ has been excluded from the integrand. $R_u = \sqrt{\frac{A_{cell}}{\pi}}$. $J_1$ is the first-order Bessel function. In general, the light source is partially coherent, i.e. it has a finite coherence length, $\lambda_c$. To include the effect of partially coherent light source, we replace the structure factor in Eq. (10) by

$$\begin{aligned} S_1(\mathbf{K}_n) &= \sum_{j \neq 1} e^{-i\mathbf{K}_n \cdot (\mathbf{R}_j - \mathbf{R}_1)} e^{-(\mathbf{R}_j - \mathbf{R}_1)^2 / 2\lambda_c^2} \\ &\approx \frac{2\pi\lambda_c^2}{A_{cell}} e^{-K_n^2 \lambda_c^2 / 2} - \frac{2\pi}{A_{cell}} \int_0^{R_u} J_0(K_n R) e^{-R^2 / 2\lambda_c^2} R dR \end{aligned}. \tag{11}$$

The coherence length $\lambda_c$ can be treated as a fitting parameter, and we found that taking $\lambda_c$ equal to 3500 nm gives an over-all good fit for all experimental spectra obtained. [30]

The advantage of using the Green's function approach is that one only needs to solve for the local function $\mathbf{u}(\mathbf{r})$ for $\mathbf{r}$ inside the nanoparticle, where the perturbation $V_1(\mathbf{r})$ is non-zero. Then, the perturbed wave function anywhere in space can be calculated by using the L-S equation, once $\mathbf{u}(\mathbf{r})$ is known inside the nanosphere. To solve for $\mathbf{u}(\mathbf{r})$ inside the nanosphere numerically, we expand each component of the vector wave function $\mathbf{u}(\mathbf{r})$ in terms of the following basis functions: $\Phi_{\ell m}(\mathbf{r}) \equiv j_\ell(k_1 r) Y_{\ell m}(\theta, \phi)$, where $j_\ell$ is the spherical Bessel function of order $\ell$ and $Y_{\ell m}$ is the spherical harmonic function with quantum numbers $(\ell, m)$. Note that all of these basis functions satisfy the Maxwell's equations for a uniform material with dielectric constant $\varepsilon_a$ (here for Au) if we choose $k_1^2 = \varepsilon_a (\omega/c)^2$. Since $\varepsilon_a$ is a complex number, so is $k_1$. The basis set $\Phi_{\ell m}$ defined this way is a complete set for finding the wave function inside the nanosphere. Within this basis set (with a cut-off at some maximum value of $\ell$), Eq. (5) can be solved efficiently via either a direct solver or an iterative solver such as the quasi-minimum residue (QMR) method. Note that the number of spherical harmonics basis functions needed to give accurate solution is much less than that used in [20, 22], in which cylindrical basis functions of the form $B_i(z)\rho^\nu e^{im\phi}$ were used. The cylindrical basis functions adopted in [20, 22] are suitable for any object with cylindrical symmetry. However, when applied to spherical objects of interest here, they are not as good as the spherical harmonics basis, since the number of basis functions needed to describe the accurate solution would be much larger and it would take much longer computation time to obtain convergent results, especially for wavelengths near the plasmonic resonance.

In order to describe the contribution of clustering nanoparticles, we introduce three types of local functions: $\mathbf{u}_1(\mathbf{r})$ for uncluttered nanoparticles, $\mathbf{c}_\alpha(\mathbf{r})$ for small nanoclusters (each of which covers an area less than $A_{cell}$), and $\mathbf{p}(\mathbf{r})$ for nanoparticles appearing in patches of aggregated nanoparticles (each of which has a size larger than $A_{cell}$ and comparable to the coherence length). $N_u, N_c$ and $N_p$ denote the numbers of cells occupied by uncluttered nanoparticles, nanoclusters, and patches, respectively. The sum $N = N_u + N_c + N_p$ represents the total numbers of cells in the area of interest. The fractions of areas occupied by clusters, patches, and uncluttered nanoparticles are denoted by $f_c = N_c/N$, $f_p = N_p/N$, and

$f_u = N_u/N = 1 - f_c - f_p$. The L-S equation for the local function of a uncluttered nanoparticle at site 1 (with $R_1 = 0$) becomes

$$\mathbf{u}_1(\mathbf{r}) = \sqrt{N}\mathbf{E}_0(\mathbf{r}) + \int \frac{d\phi_n}{(2\pi)^2} \int k_n dk_n \int d\mathbf{r}' e^{i\mathbf{k}_n \cdot (\mathbf{\rho}-\mathbf{\rho}')} \mathbf{g}_n(z,z') \cdot \left\{ \left[1 + S_1(K_n) f_u f\right] V_1(\mathbf{r}') \mathbf{u}_1(\mathbf{r}') \right.$$
$$\left. + \sum_\alpha p_\alpha f S_1(K_n) V_\alpha(\mathbf{r}') \mathbf{c}_\alpha(\mathbf{r}') \right\}$$

(12)

where $V_\alpha(\mathbf{r}') = (\varepsilon_a - 1)(\omega/c)^2$ for $\mathbf{r}'$ inside a cluster of type $\alpha$ and vanishes otherwise. $p_\alpha$ denotes the weighting factor for clusters of type $\alpha$ with $\sum_\alpha p_\alpha \eta_\alpha = f_c$, where $\eta_\alpha = A_\alpha / A_{cell}$ with $A_\alpha$ denotes the area covered by a single cluster of type $\alpha$. To avoid using too many fitting parameters, we assume that $p_\alpha$ is inversely proportional to the area occupied by the cluster, i.e. $p_\alpha = p_0 / \eta_\alpha$ with $p_0 = f_c / n_c$, where $n_c$ is the number of different sizes of clusters included. Note that $f_c$ (total fraction of small clusters) is typically less than 5% for the samples studied; thus, the last term describing the effect of surrounding clusters on the local function $\mathbf{u}_1(\mathbf{r})$ is negligible. However, the influence of $\mathbf{u}_1(\mathbf{r})$ on $\mathbf{c}_\alpha(\mathbf{r})$ is non-negligible as we shall see below.

Within the same approximation, the local function $\mathbf{c}_\alpha$ for a cluster of type $\alpha$ satisfies a similar L-S equation

$$\mathbf{c}_\alpha(\mathbf{r}) = \sqrt{N}\mathbf{E}_0(\mathbf{r}) + \int \frac{d\phi_n}{(2\pi)^2} \int k_n dk_n d\mathbf{r}' e^{i\mathbf{k}_n \cdot (\mathbf{\rho}-\mathbf{\rho}')} \mathbf{g}_n(z,z') \cdot \left\{ \left[1 + S_1(K_n) p_\alpha f\right] V_\alpha(\mathbf{r}') \mathbf{c}_\alpha(\mathbf{r}') \right.$$
$$\left. + \sum_{\alpha' \neq \alpha} S_1(K_n) p_{\alpha'} f V_{\alpha'}(\mathbf{r}') \mathbf{c}_{\alpha'}(\mathbf{r}') + f_u f S_1(K_n) V_1(\mathbf{r}') \mathbf{u}_1(\mathbf{r}') \right\}$$

(13)

for $\mathbf{r} \in C_\alpha$, where $C_\alpha$ denotes a local domain surrounding a cluster of type $\alpha$. The last term describes the coupling of the cluster to surrounding nanoparticles, which is non-negligible since $f_u f$ is close to 1. For simplicity, we can approximate the average of all clusters of type $\alpha$ by a spheroid with height $h$ and diameter $d_\alpha$ for the horizontal cross-section, as depicted in Fig. 2(b). That is, we use a pancake-like oblate spheroid [see discs outlined by a dash line in Fig. 2(b)] to describe the average of these clusters. For example, the angular average of close-packed clusters made of two or three nanoparticles may be represented by a pancake with diameter $2d$, which occupies the volume covered by a circular revolution of a dimer. Similarly, the angular average of closed-packed clusters made of five to seven nanoparticles may be simulated by a pancake with diameter $3d$, which occupies the volume covered by a circular revolution of a chain of three nanoparticles, and the angular average of clusters made of eight to twelve nanoparticles is simulated by a pancake with diameter $4d$. For not so closely packed cluster of 3-12 particles, we may approximate them by pancakes with diameters between $2d$ and $4d$. Thus, a model including pancakes with diameters evenly distributed between $2d$ and $2R_u$ (the cut-off diameter) with suitable proportions $p_\alpha$ can roughly describe the angular averages of all clusters which can fit into a cell with area $A_{cell}$.

The larger clusters are approximated by patches of aggregated nanoparticles, which have a close-packed structure similar to the largest patch sketched in Fig. 2(b), expect that the size can be comparable to the coherence length. The average wave function for a nanoparticle within the patch $\mathbf{p}(\mathbf{r})$ is assumed to satisfy an equation similar to Eq. (5), which is for a uniform and randomly distributed system.

$$\mathbf{p}(\mathbf{r}) = \sqrt{N} e^{-i\mathbf{k}_0 \cdot \mathbf{R}_i} \mathbf{E}_0(\mathbf{r}) + \int \frac{d\phi_n}{(2\pi)^2} \int k_n dk_n \left[1 + S_p(\mathbf{K}_n) f\right] \int d\mathbf{r}'_i e^{i\mathbf{k}_n \cdot (\boldsymbol{\rho} - \boldsymbol{\rho}')} \mathbf{g}_n(z, z') \cdot V_1(\mathbf{r}') \mathbf{p}(\mathbf{r}') \quad (14)$$

where $\mathbf{r}$ is restricted within a nanoparticle in the clustering patch. $S_p(\mathbf{K}_n)$ denotes the structure factor for nanoparticles in a clustering patch, which is similar to $S_1(\mathbf{K}_n)$ as given by Eq. (11), except that the cell area is replaced by $A_p = \sqrt{3}d^2/2$ (for two-dimensional closely-packed structure) and $R_u$ replaced by $R_p = \sqrt{A_p/\pi}$.

Here, we have neglected the coupling of $\mathbf{p}(\mathbf{r})$ with $\mathbf{u}_1(\mathbf{r})$ and $\mathbf{c}_\alpha(\mathbf{r})$, since the average separation between a patch (i.e. large cluster) and a nanoparticle (or small cluster) is larger than the coherent length in our system. To solve for the local functions $\mathbf{u}_1(\mathbf{r})$, $\mathbf{c}_\alpha(\mathbf{r})$, and $\mathbf{p}(\mathbf{r})$ for $\mathbf{r}$ inside an uncluttered nanoparticle, small clusters, or a nanoparticle in large patches, we expand them in terms of products of spherical Bessel functions $j_\ell(k_1 r)$ and spherical harmonics functions $Y_{\ell m}(\Omega)$ as described above. We noticed that for oblate spheroids with low $h/d_\alpha$ ratio, larger cut-off for the angular momentum quantum number is needed to give convergent results.

The coupled equations (12) and (13) are solved numerically to obtain the local functions $\mathbf{u}_1(\mathbf{r})$ and $\mathbf{c}_\alpha(\mathbf{r})$ and Eq. (14) is used to solve for local function $\mathbf{p}(\mathbf{r})$ in terms of spherical harmonics basis functions. Finally, the principal-order reflection coefficient is obtained by calculating the average **E**-field on a plane just above the nanoparticles (taken to be at $z=0$) according to the following equation

$$\mathbf{E}(z=0) = \mathbf{E}_0(z=0) + \frac{1}{A_{cell}} \int dz' \mathbf{g}_0(0,z') \cdot \int d\phi' \int \rho' d\rho' e^{-i\mathbf{k}_0 \cdot \boldsymbol{\rho}'} [f_u V_1(\mathbf{r}') \mathbf{u}_1(\mathbf{r}') + \sum_\alpha p_\alpha V_\alpha(\mathbf{r}') \mathbf{c}_\alpha(\mathbf{r}') + f_p V_1(\mathbf{r}') \mathbf{p}(\mathbf{r}')] .$$

(15)

The TE (*s*-polarized) and TM (*p*-polarized) reflectivities are given by $r_s = E_y / E_{oy} - 1 ; r_p = H_y / H_{oy} - 1$, respectively, from which we can obtain the ellipsometric parameters $\Psi = \tan^{-1}|r_p / r_s|$ and $\Delta = \arg(r_p / r_s)$.

### III. Results and discussions

To check the accuracy of the current SHGF method, we have compared the calculated near-field at the top of the sphere ($z=0$) for light scattering from an isolated Au sphere with those obtained by the Mie theory. The comparison is shown in Fig. 3. The solid line denotes the Mie theory result. The dash-dotted line, dotted line and dashed line denote results obtained by the SHGF method with various cutoff values of $\ell$ and number of $k_n$ points : ($\ell_{max}$, $N_k$) = (2,101), (4,51) and (4,101), respectively. The numbers of $z$ and $\rho$ meshes used are both 50 in these cases. When we consider a sphere with a larger diameter, larger values for ($\ell_{max}$, $N_k$) are needed to give convergent results. For ($\ell_{max}$, $N_k$) = (4,101), we find convergent results (dashed lines) for all cases, and they are in excellent agreement with the Mie theory, indicating the accuracy of our numerical implementation.

Next, we compare our theoretical results (without and with clustering effects) with the SE measurements. In order to ensure convergent results, we have used a cutoff of angular momentum quantum number, $\ell_{max} = 6$ for nanoparticles and $\ell_{max} = 6$ with doubled number of $z$ meshes for the pancake like clusters due to their low aspect ratio of height ($h$) to diameter ($d_\alpha$). We found that the minimum value of $\ell_{max}$ needed to get reasonable results is 3. For the model without clusters, the fitting parameters used include: $d$ (diameter) and $h$ (height) of nanoparticles, the average distance between nanoparticles (pitch), $p = \sqrt{A_{cell}}$, and the similarity factor *f*. For the model including coupling with clusters, we adopt the best-fit parameters from the model without clusters and add two fitting parameters: the total fraction of area occupied by small clusters $f_c$ and that for the patches of aggregated nanoparticles, $f_p$. Seven small clusters ($n_c = 7$) with diameters evenly distributed between $2d$ and $2R_u$ are used, and their heights are kept the same as the nanoparticles. We find that the results remain nearly unchanged even if we increase the sampling number of sizes for small clusters.

Figure 4 shows the ellipsometric parameters, $\Psi$ and $\Delta$ as functions of photon energy obtained by the SHGF method without clustering effect and experimental data for four different sizes of nanoparticles with nominal diameters of 20nm, 40nm, 60nm and 80nm, at three different angles of incidence: 55, 60 and 65 degrees. The parameters used to get the best-fit ellipsometric spectra are listed in Table 1. We find that the theoretical results agree fairy well with SE measurements for the 20nm case. For samples with larger size nanoparticles, the fit is not as good, especially for photon energies less than 2.5 eV, where the plasmonic resonance dominates. There is aslo significant deviation from data for $\Psi$ for photon energies larger than 5eV. [See Fig. 4 (b)-(d)] The present theoretical predictions based on the SHGF method are in better agreement (especially for the 80nm case) with experimental data than previous calculations reported in Ref. [22], which were based on the FEGF method with the use of cylindrical basis functions. The improvement is mainly due to the better basis functions used (which gives better results near the plasmonic resonance) and the inclusion of effect due to finite coherent length (which improves the over-all spectrum shape for photon energies larger than 5eV). The computation time is also improved (CPU time spent to compute $\Psi$ and $\Delta$ for each photon energy at a given angle of incidence is about 2.5 or 6 seconds for $\ell_{max} = 3$ or 5 on a single Intel 1.83GHz processor)

Figure 5 shows the calculated ellipsometric parameters, $\Psi$ and $\Delta$ as functions of photon energy obtained by the SHGF method for four different isolated pancakes of diameters $2d$, $2.5d$, $3d$, and $3.5d$ with $d = 60$ nm at three different angles of incidence: $55^o$ (solid lines), $60^o$ (dashed lines) and $65^o$ (dash-dotted lines). The number of $z$ meshes used in the integration is kept at 50 in all cases and the pitch ($p$) used is 170 nm. In order to ensure the convergent results, the cutoff of angular momentum quantum number $\ell$ ($\ell_{max}$), the number of $\rho$ meshes ($N_\rho$), and the number of $k_n$ meshes ($N_k$) used in the integration for pancakes with lateral sizes $2d$, $2.5d$, $3d$, and $3.5d$ are ($\ell_{max}$, $N_\rho$, $N_k$) = (6,70,71), (6,80,81), (6,90,91), and (7,100,91), respectively. As can be seen in Fig. 5, these clusters represented by pancakes lead to larger enhancement and multiple plasmonic resonances at energies near 2 eV due to their enlarged cross-section in the x-y plane.

Figure 6 shows the calculated ellipsometric spectra (for $\Psi$ and $\Delta$), taking into account the clustering effect of nanoparticles. The additional parameters $f_c$ and $f_p$ used to obtain the best fit are listed in Columns 5 and 6 of Table 1. A variation of $f_c$ and $f_p$ by more than ±0.01 will lead to significantly worse fit. For this calculation, the similarity factor $f$ is rescaled such that $f_u f$ is the same as the $f$ listed in Column 3 of Table 1. Including the clustering effect, we can obtain much better agreement with the experimental data for samples with nominal sizes of 40nm, 60nm, and 80nm, where there is obvious cluster formation as shown in Fig. 1(b)-(d). The mean-squared errors (MSE's) as shown in Column 6 of Table 1 now become much smaller. Examining the contributions, we found that the coupling with smaller clusters are responsible for the improvement for photon energies below 2.5 eV, where the plasmonic resonance plays an important role. Comparing the spectra in Fig. 6 and Fig. 5 for photon energies below 2.5 eV, we notice a qualitative difference in the spectral lineshapes with and without the clustering effect. Without the clustering effect, the $\Psi$ spectra display a simple peak structure near 2.5 eV, which reflects the plasmonic resonance of the Au nanoparticles. With the presence of clustering effect, the $\Psi$ spectra display asymmetric peak structure with a dip near 1.5eV and a broad peak covering 2 eV to 2.5 eV, indicating effects due to coupling of multiple plasomonic resonances. All these features can be seen in the modeling results when the pancake like clusters are included and they are in reasonable agreement with the experimental data. Furthermore, for photon energies above 5 eV, the agreement with experimental data for the $\Psi$ spectra is improved considerably when we also mix in the contribution from the patches of closely packed nanoparticles as can be seen in Fig. 6.

## IV. Conclusion

In this paper, we have presented a spherical harmonics-based Green's function (SHGF) method for studying randomly distributed nanoparticles on multilayer films, taking into account the clustering effect. With the use of spherical harmonics as basis functions, the light scattering from such a system can be handled efficiently and accurately. This improved technique allows us to analyze complicated models including the coupling of pancake-like nanoclusters with the randomly distributed nanospheres. Our modeling results including coupling with clusters agree well with the spectroscopic ellipsometry data. By comparing the theoretical predictions with experimental measurements based on spectroscopic ellipsometry, we can provide structure information of the distribution of nanoparticles on a substrate, including the average spacing of nanoparticles ($p$), the fraction of areas occupied by small clusters ($f_c$) and that of patches of closely-packed nanopaticles ($f_p$). Such information is very useful for nondestructive metrology of nanoparitcles covered samples.

**Acknowledgements**

This work was supported in part by the Nanoproject of Academia Sinica and National Science Council of the Republic of China under Contract No. NSC 101-2112-M-001-024-MY3.

# References

[1] H. Raether, *Surface Plasmons on Smooth and Rough Surfaces and on Gratings. Springer Tracts in Modern Physics*, Vol. 111, (New York, Springer-Verlag, 1988).

[2] J. T. Krug, G. D. Wang, S. R. Emory and S. Nie, "Efficient Raman enhancement and intermittent light emission observed in single gold nanocrystals," J. Am. Chem. Soc. **121**, 9208-9214 (1999).

[3] H. Xu, J. Aizpurua, M. Käll and P. Apell, "Electromagnetic contributions to single-molecule sensitivity in surface-enhanced Raman scattering," Phys. Rev. E **62**, 4318-4324 (2000).

[4] A. M. Michaels, J. Jiang and L. Brus, "Ag nanocrystal junctions as the site for surface-enhanced Raman scattering of single rhodamine 6G molecules," J. Phys. Chem. B **104**, 11965-11971 (2000).

[5] A. Wokaun, J. P. Gordon and P. F. Liao, "Radiation Damping in Surface-Enhanced Raman Scattering," Phys. Rev. Lett. **48**, 957-960 (1982).

[6] S. M. Nie and S. R. Emery, "Probing single molecules and single nanoparticles by surface-enhanced Raman scattering," Science **275**, 1102-1106 (1997).

[7] S. V. Gaponenko, A. A. Gaiduk, O. S. Kulakovich, S. A. Maskevich, N. D. Strekal and V. M. Shelekhina, "Raman scattering enhancement using crystallographic surface of a colloidal crystal," JETP Letters **74**, 309-311 (2001).

[8] B. Kaplan, T. Novikova, A. D. Martino and B. Drévillon, "Characterization of bidimensional gratings by spectroscopic ellipsometry and angle-resolved Mueller polarimetry," Appl. Opt. **43**, 1233-1240 (2004).

[9] H. Wormeester, E. S. Kooij, A. Mewe, S. Rekveld and B. Poelsema, "Ellipsometric characterisation of heterogeneous 2D layers," Thin Solid Films **455-456**, 323-334 (2004).

[10] S. H. Hsu, E. S. Liu, Y. C. Chang, J. N. Hilfiker, Y. D. Kim, T. J. Kim, C. J. Lin and G. R. Lin, "Characterization of Si nanorods by spectroscopic ellipsometry with efficient theoretical modeling," Phys. Status Solidi A **205**, 876-879 (2008).

[11] D. Schmidt, B. Booso, T. Hofmann, E. Schubert, A. Sarangan and M. Schubert, "Monoclinic optical constants, birefringence, and dichroism of slanted titanium nanocolumns determined by generalized ellipsometry," Appl. Phys. Lett. **94**, 011914 (2009).

[12] G. Mie, "Beiträge zur optik trüber medien, speziell kolloidaler metallösungen," Ann. Phys. **330**, 377-445 (1908).

[13] C. F. Bohren and D. R. Huffman, *Absorption and Scattering of Light by Small Particles* (Wiley, New York, 1983).

[14] S. Asano and G. Yamamoto, "Light scattering by a spheroidal particle," Appl. Opt. **14**, 29-49 (1975).

[15] G. Videen, "Light scattering from a sphere on or near a surface," J. Opt. Soc. Am. A **8**, 483-489 (1991).

[16] I. Simensen, R. Lazzari, J. Jupille and S. Roux, "Numerical modeling ot the optical response of supported metallic particles," Phys. Rev. B **61**, 7722-7733 (2001).

[17] R. Lazzari, I. Simensen, D. Bedeaux, J. Vlieger and J. Jupille, "Polarizability of truncated spheroidal particles supported by a substrate: model and applications," Eur. Phys. J. B **24**, 267-284 (2001).

[18] D. Bedeaux and J. Vlieger, *Optical Properties of Surfaces* (Imperial College Press, London, UK, 2002).

[19] M. G. Moharam, E. B. Grann, D. A. Pommet and T. K. Gaylord, "Formulation for stable and efficient implementation of the rigorous coupled-wave analysis of binary gratings," J. Opt. Soc. Am. A **12**, 1068-1076 (1995).

[20] Y. C. Chang, S. H. Hsu, P. K. Wei and Y. D. Kim, "Optical nanometrology of Au nanoparticles on a multilayer film," Phys. Status Solidi C **5**, 1194–1197 (2008).

[21] Y. C. Chang, G. Li, H. Chu and J. Opsal, "Efficient finite-element, Green's function approach for critical dimension metrology of three-dimensional gratings on multilayer films," J. Opt. Soc. Am. A **23**, 638-645 (2006).

[22] S. H. Hsu, Y. C. Chang, Y. C. Chen, P. K. Wei and Y. D. Kim, "Optical metrology of randomly-distributed Au colloids on a multilayer film," Opt. Exp. **18**, 1310-1315 (2010).

[23] R. Lazzari and I. Simonsen, "GRANFILM: a software for calculating thin-layer dielectric properties and Fresnel coefficients," Thin Solid Films **419,** 124-136 (2002).

[24] G. R. Lin, Y. C. Chang, E. S. Liu, H. C. Kuo and H. S. Lin, "Low refractive index Si nanopillars on Si substrate, Appl. Phys. Lett. **90**, 181923 (2007).

[25] R. S. Moirangthema, Y. C. Chang and P. K. Wei, "Investigation of surface plasmon biosensing using gold nanoparticles enhanced ellipsometry," Opt. Lett. **36**, 775 (2011).

[26] R. S. Moirangthema, Y. C. Chang and P. K. Wei, "Ellipsometry study on gold-nanoparticle-coated gold thin film for biosensing application," Bio. Opt. Exp. **2**, 2569 (2011).

[27] E. D. Palik, ed., *Handbook of Optical Constants of Solids*, vol. 1 (Academic, Orlando, FL, USA, 1985).

[28] H. Y. Xie, Y. C. Chang, G. Li and S. H. Hsu, "Effect of clustering on ellipsometric spectra of randomly distributed gold nanoparticles on a substrate",

Opt. Exp., **21**, 3091–3102 (2013).

[29] See for example, Fayyazuddin and Riazuddin, Quantum Mechanics (World Scientific, 1990), p. 368.

[30] Here we correct an error of Ref. [28], in which Eq. (11) should be replaced by Eq. (11) in the present paper. With this correction and by using a larger (convergent) set of basis functions, the best-fit value for the coherent length becomes 3500nm.

**Table and Figure caption**

**Table:**

[1] Best-fit parameters used in the theoretical modeling for Au nanoparticles without and with clusters.

**Figure:**

[1] SEM pictures of samples with random distribution of Au nanoparticles with diameters: (a) 20nm, (b) 40nm, (c) 60nm, and (d) 80nm.

[2] (a) a picture which describes randomly distributed identical nanoparticles with variable distances between the centers of particles $R_i$ and the origin $O$ (top view); (b) equivalent spheroid model to describe clusters with three different diameters (top view).

[3] The calculate field strength, $|\mathbf{E}|$ at the top of the sphere as a function of photon energy for light scattering from an isolated Au sphere obtained by the present Green's function method with different cutoffs: ($\ell_{max}$, $N_k$) = (2,101) (dash-dotted), (4,51) (dotted), and (4,101) (dashed) and Mie scattering theory (solid). (a) $d = 20$ nm, (b) $d = 40$ nm, (c) $d = 60$ nm, and (d) $d = 80$ nm.

[4] SE measurements (solid curves) and model calculations (dash-dotted curves) without clusters for random distribution of nanoparticles with nominal sizes of (a) 20 nm, (b) 40 nm, (c) 60 nm, and (d) 80 nm for incident angles of 55°, 60°, and 65°.

[5] The ellipsometric parameters, $\Psi$ and $\Delta$ as functions of photon energy obtained by the SHGF method for four different isolated pancakes of diameters (a) $2d$, (b) $2.5d$, (c) $3d$ and (d) $3.5d$ with $d = 60$nm at three different angles of incidence: 55° (solid line), 60° (dashed line) and 65° (dash-dotted line) on the substrate.

[6] SE measurements (solid curves) and model calculations (dashed dot curves) including clusters for random distribution of nanoparticles with nominal sizes of (a) 40, (b) 60, and (c) 80 nm for incident angles of 55°, 60°, and 65°.

| Nanoparticle height $h$ (nm) | Aspect ratio ($h/d$) | Similarity factor, $f$ | Average pitch, $p$ (nm) | Fraction of small clusters, $f_c$ | Fraction of nanoparticles in patches, $f_p$ | MSE for noncluster model | MSE for cluster model |
|---|---|---|---|---|---|---|---|
| 18 | 0.95 | 1.0 | 50 | | | 5.50 | |
| 38 | 0.90 | 0.8 | 140 | 0.015 | 0.005 | 8.00 | 5.60 |
| 66 | 0.98 | 0.7 | 170 | 0.02 | 0.03 | 9.94 | 6.36 |
| 84 | 0.97 | 0.7 | 220 | 0.025 | 0.03 | 11.46 | 6.55 |

**Table. 1**

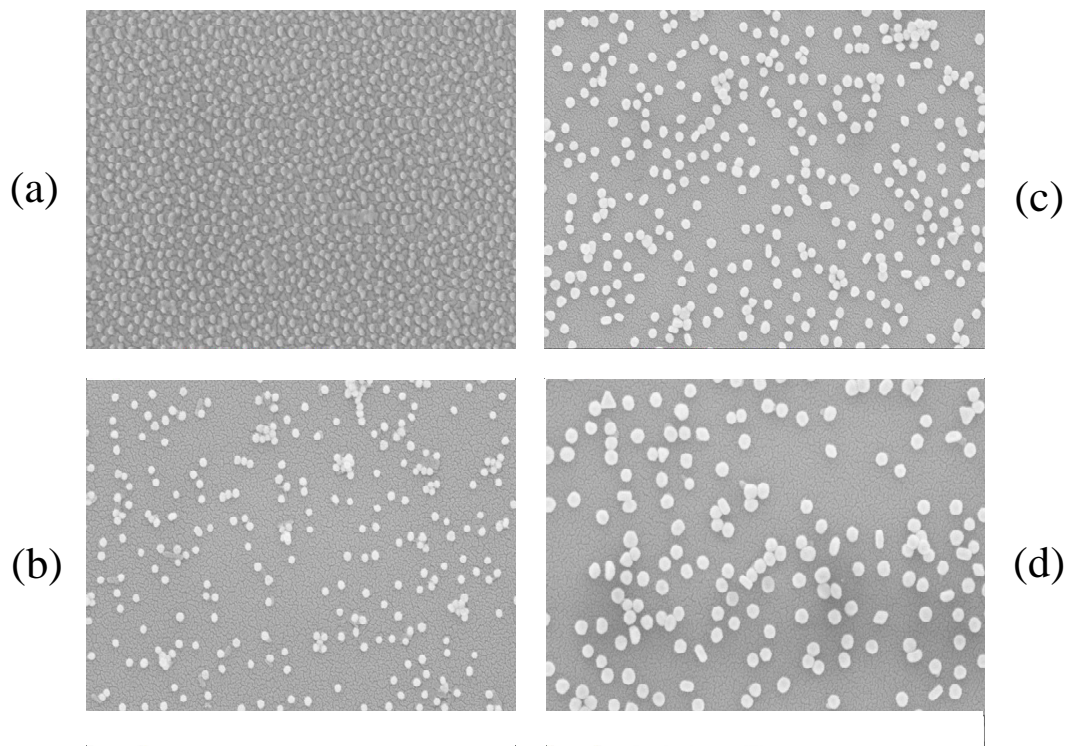

**Fig. 1**

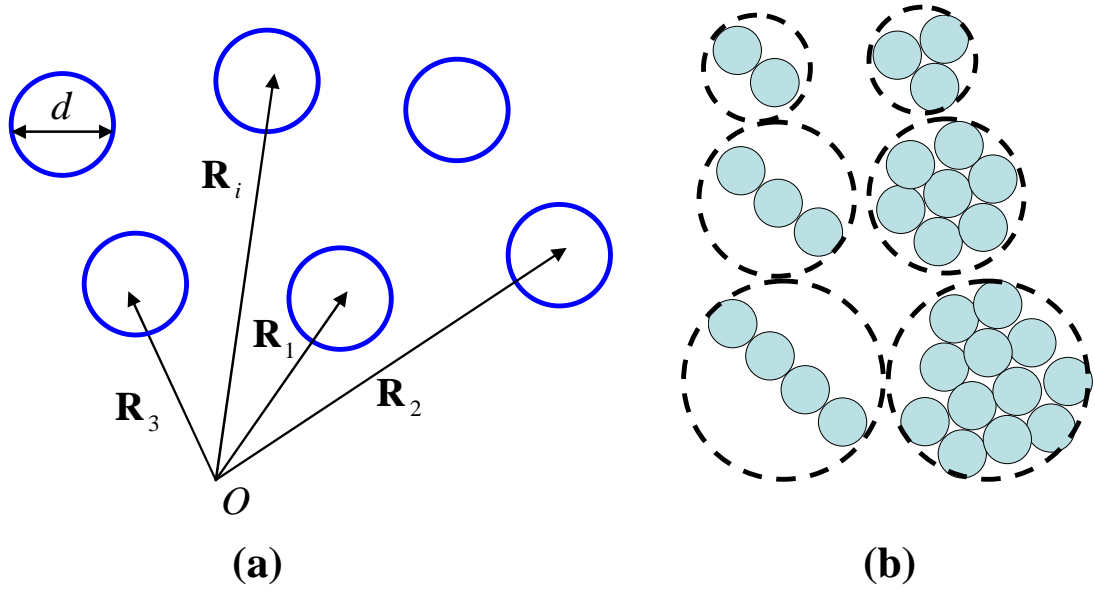

**Fig. 2**

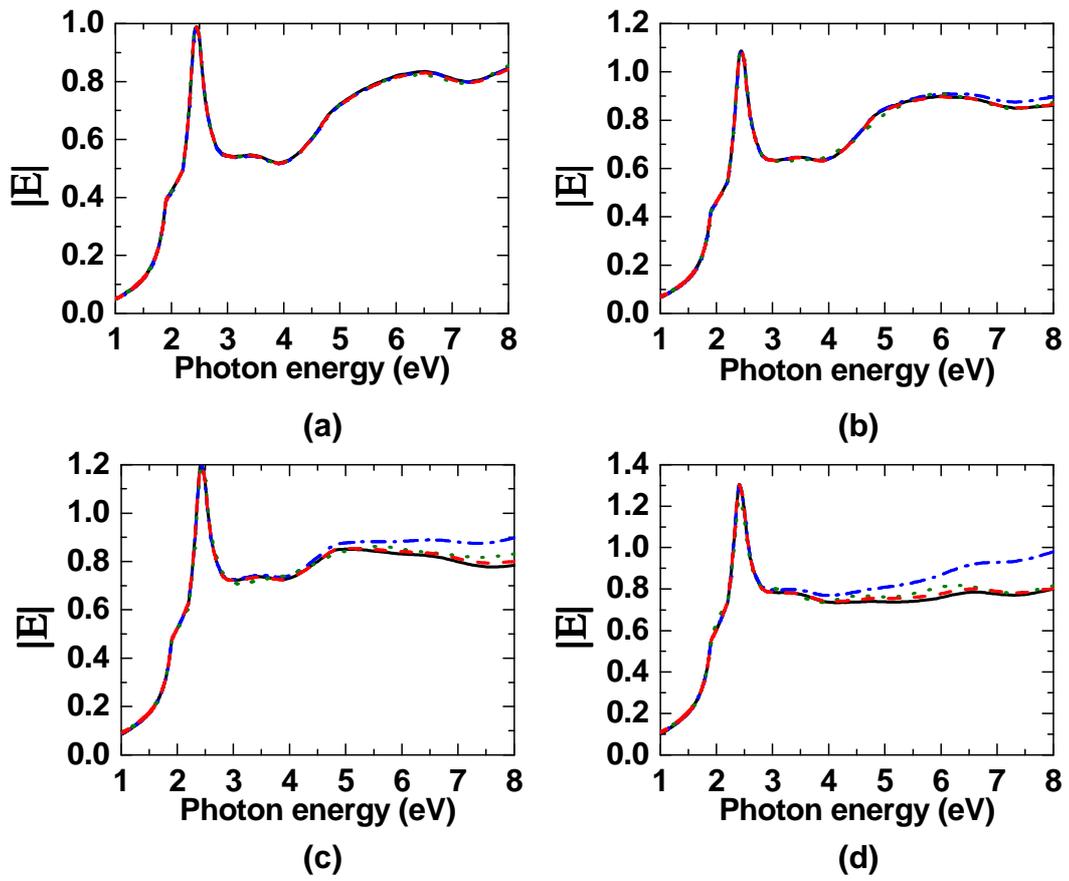

Fig. 3

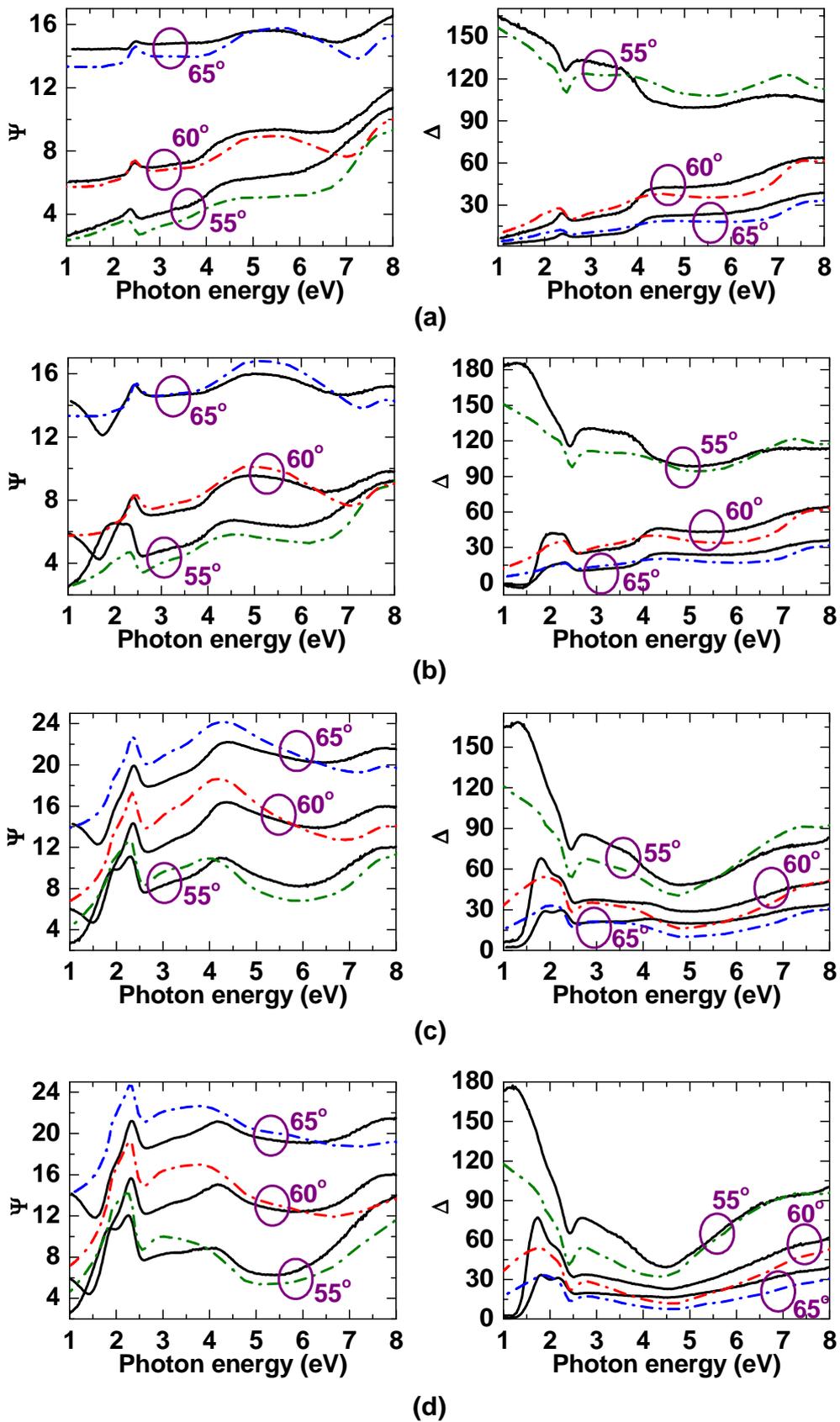

Fig. 4

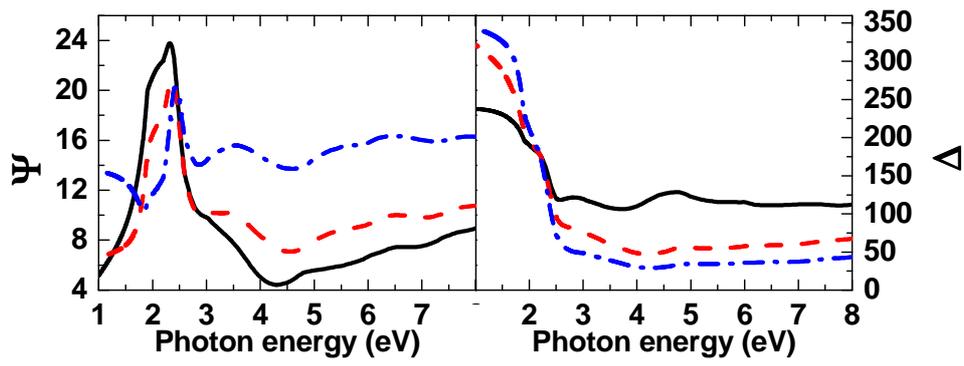

(a)

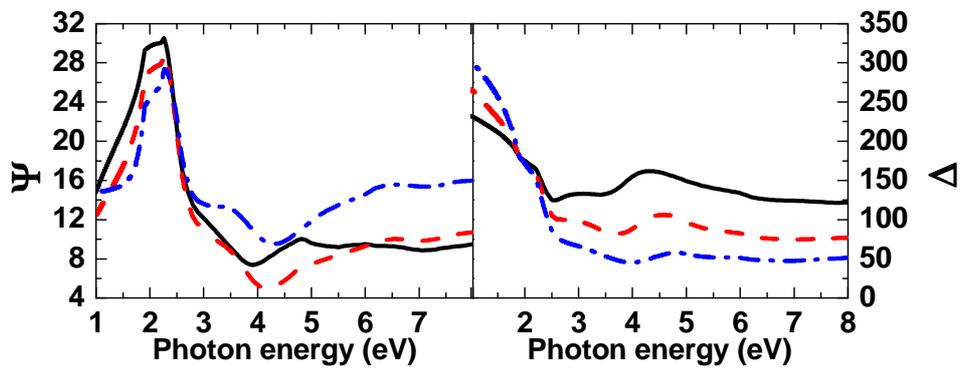

(b)

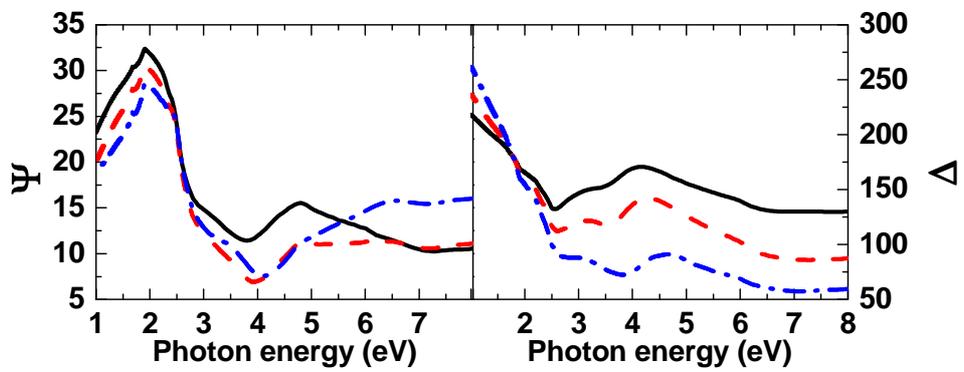

(c)

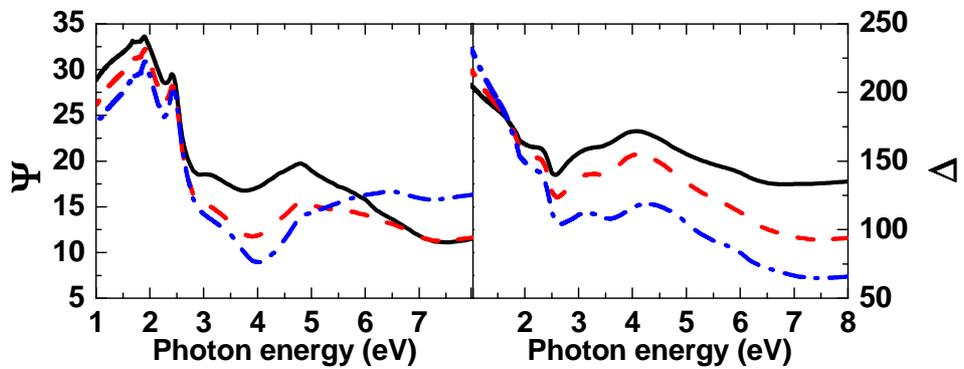

(d)

**Fig. 5**

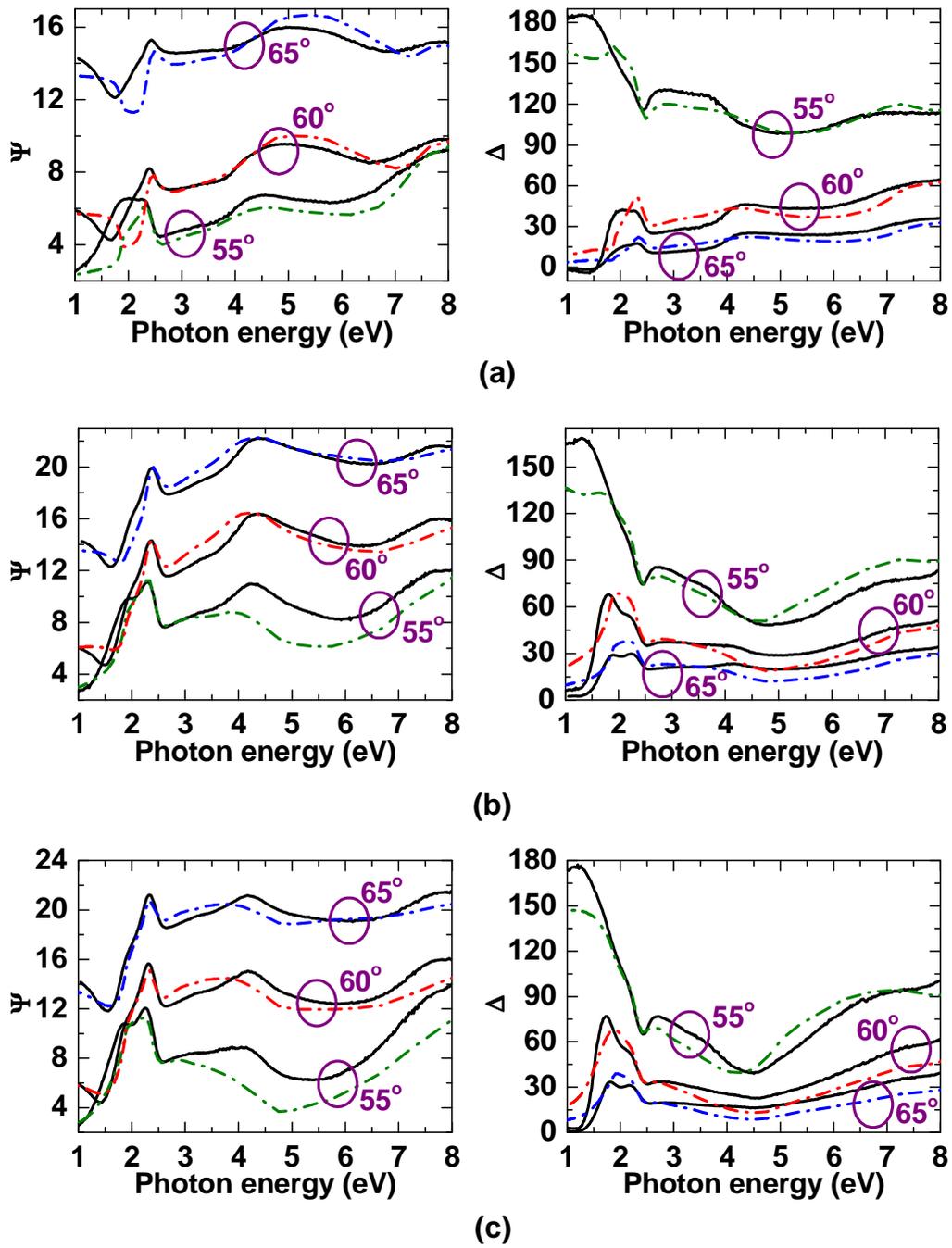

**Fig. 6**